\documentclass{article}
\usepackage{spconf,amsmath,graphicx,hyperref}
\usepackage{booktabs}
\usepackage{graphicx}
\usepackage{multirow}
\usepackage{todonotes}
\usepackage[table]{xcolor}
\usepackage{enumitem}
\usepackage{placeins}

\title{Fed-PISA: Federated Voice Cloning via Personalized Identity-Style Adaptation}
\name{Qi Wang\textsuperscript{1,2,4}, Shituo Ma\textsuperscript{3,4}, Guoxin Yu\textsuperscript{1}, Hanyang Peng\textsuperscript{1}, Yue Yu\textsuperscript{1}}
\address{  \textsuperscript{1}Peng Cheng Laboratory \ \ \ \   \textsuperscript{2} Institute of Computing Technology, Chinese Academy of Sciences
 \\
  \textsuperscript{3}Institute of Information Engineering, Chinese Academy of Sciences
 \\
 \textsuperscript{4}University of Chinese Academy of Sciences
 \\
 }
\begin{document}
\ninept
\maketitle
\begin{abstract}

Voice cloning for Text-to-Speech (TTS) aims to generate expressive and personalized speech from text using limited data from a target speaker. Federated Learning (FL) offers a collaborative and privacy-preserving framework for this task, but existing approaches suffer from high communication costs and tend to suppress \emph{stylistic heterogeneity}, resulting in insufficient personalization. To address these issues, we propose Fed-PISA, which stands for \textbf{Fed}erated \textbf{P}ersonalized \textbf{I}dentity-\textbf{S}tyle \textbf{A}daptation. To minimize communication costs, Fed-PISA introduces a disentangled Low-Rank Adaptation (LoRA) mechanism: the speaker's timbre is retained locally through a private ID-LoRA, while only a lightweight style-LoRA is transmitted to the server, thereby minimizing parameter exchange. To harness heterogeneity, our aggregation method, inspired by collaborative filtering, is introduced to create custom models for each client by learning from stylistically similar peers. Experiments show that Fed-PISA improves style expressivity, naturalness, and speaker similarity, outperforming standard federated baselines with minimal communication costs.

\end{abstract}

\section{Introduction}
Modern Text-to-Speech (TTS) systems~\cite{du2025cosyvoice, le2023voicebox, xie2025fireredtts} increasingly leverage voice cloning—the task of synthesizing speech in a target speaker's voice from arbitrary text inputs— to enable deep personalization: the synthesis of speech that captures not only a target speaker's vocal timbre but also their unique prosodic patterns and expressive style. The technical pursuit of this nuanced personalization has motivated the emergence of distinct research paradigms.

\begin{figure*}[!t]
    \centering
    \includegraphics[width=0.82\linewidth]{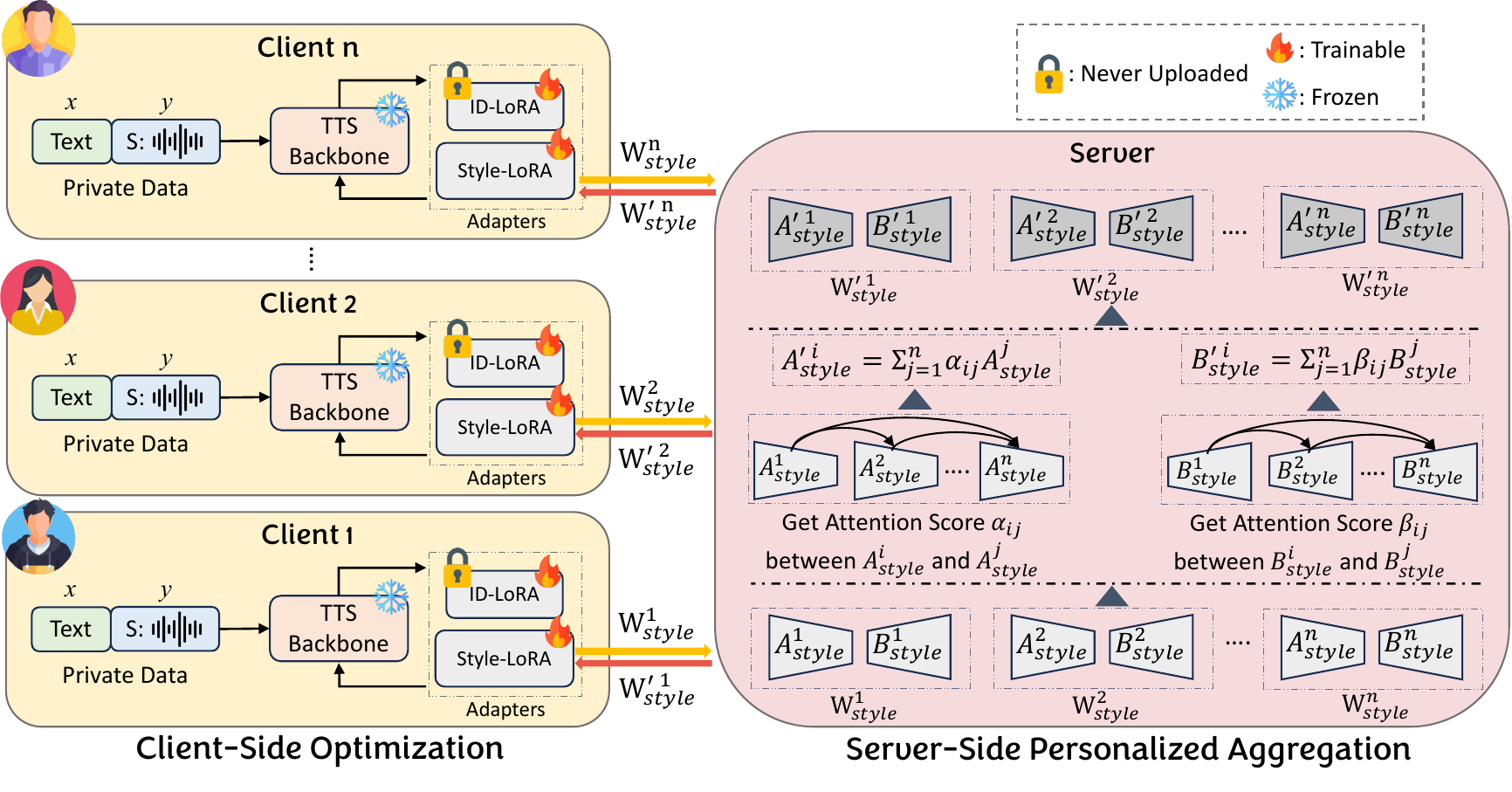}
    \caption{An overview of the FED-PISA framework. On the client side, a private ID-LoRA captures speaker timbre locally, while only a lightweight Style-LoRA is trained and uploaded for aggregation. The server then employs a personalized aggregation strategy to create a custom style model for each client by learning from stylistically similar peers.}
    \label{fig:fig_overview}
\end{figure*}

Among these, two main directions have emerged: high-fidelity zero-shot personalization based on style vectors or large-scale corpus~\cite{du2025cosyvoice, le2023voicebox, meng2025ds, chen2024emoknob}, along with lightweight on-device deployment via model pruning and Parameter-Efficient Fine-Tuning (PEFT)~\cite{huang2023personalized, comini2025lightweight, bondaruk2025lorp, kwon2025parameter, fujita2024lightweight}. 
Despite improvements in speaker similarity and efficiency, these methods share a fundamental limitation: the personalization process is isolated. 
This isolation creates ``style silos", where a client's model is constrained by their own limited data, unable to leverage the rich stylistic heterogeneity present across a wider community of peers. 
Consequently, there is an increasing need for collaborative style learning across different clients under strict preservation of the local data privacy.

Under such circumstances, Federated Learning (FL)~\cite{zhang2021survey} offers a privacy-preserving approach that allows multiple clients to collaboratively train a model while keeping their private data on local devices. Existing FL-based TTS frameworks, such as FedSpeech~\cite{jiang2021fedspeech} and Federated Dynamic Transformer~\cite{hong2021federated}, primarily focus on backbone modification or partitioning to balance speaker-specific personalization with shared global knowledge. Nevertheless, training personalized TTS systems in this manner faces two key challenges. First, these methods often incur substantial computational and communication costs due to complex backbone modifications or large-scale parameter exchanges, hindering their deployment on resource-constrained edge devices. Second, they suppress the severe \textit{stylistic heterogeneity} exhibited in speech data distribution (e.g., variations in emotion, prosody) to preserve individual speaker timbre. FedSpeech isolates style features locally by parameter masking, while Federated Dynamic Transformer retains personalized layers that capture styles locally, aggregating only the general model. Such suppression limits the model from effectively learn diverse and expressive styles across heterogeneous client data, leading to insufficient model personalization capabilities. This motivates the exploration: 
\emph{\textbf{Can a federated TTS system effectively learn and utilize \textit{heterogeneous styles} across clients while maintaining low communication costs?}}

To tackle the aforementioned issues, we propose an efficient and personalized federated TTS framework, namely \textbf{Fed}erated \textbf{P}ersonalized \textbf{I}dentity-\textbf{S}tyle \textbf{A}daptation~(\textsc{Fed-PISA}). 
\textsc{Fed-PISA} builds upon the PEFT techniques, specifically inspired by recent studies that utilize Low-Rank Adaptation (LoRA) for single-client TTS personalization~\cite{bondaruk2025lorp, kwon2025parameter, fujita2024lightweight, lou2024stylespeech}.
It adapts the LoRA to a more complex federated setting and consists of two key components: 
(i) To improve efficiency and preserve speaker timbre, a decoupled LoRA mechanism is introduced. A private ID-LoRA is trained for each client and locally frozen to capture their unique speaker timbre robustly. A lightweight Style-LoRA is used for communication on the server, effectively reducing costs. (ii) To leverage stylistic heterogeneity, we introduce a personalized aggregation strategy. Inspired by collaborative filtering in recommendation systems~\cite{isinkaye2015recommendation}, this strategy creates a custom-aggregated style model for each client by prioritizing updates from stylistically similar peers. 

Our main contributions are threefold\footnote{Codes and Audio samples are available at \url{https://huggingface.co/spaces/sDuoluoluos/FedPISA-Demo}}
:
\begin{itemize}[leftmargin=1em, itemsep=0.5ex, topsep=0.5ex]
    \item Our work addresses a key challenge overlooked by prior federated TTS frameworks: tackling high communication costs and overcoming the suppression of \textit{stylistic heterogeneity}.
    \item We propose \textsc{Fed-PISA}, an efficient framework that disentangles speaker timbre from collaborative style learning using LoRA and a novel personalized aggregation mechanism.
    \item Experiments on four public datasets demonstrate that \textsc{Fed-PISA} consistently improves style expressivity, speaker similarity,  naturalness, and communication efficiency over federated and nonfederated baselines.
\end{itemize}

\section{Methods}
\label{sec:method}


\subsection{Backbone and Adapters}
Let $x$ be the input text and $y$ the target speech waveform. During training, a frozen speech tokenizer $E(\cdot)$ converts $y$ into discrete acoustic/semantic units $u = E(y)$. We adopt \textsc{GPT-SoVITS-v4}\footnote{\url{https://github.com/RVC-Boss/GPT-SoVITS}} as a two-stage backbone: a GPT-style autoregressive semantic predictor that models $p(u\,|\,x, r)$ conditioned on text $x$ and reference audio $r$ for speaker/style cues and a SoVITS Conditional Flow Matching (CFM) decoder that renders waveform by modeling $p(y\,|\,u, r)$.

\textbf{LoRA Parameterization.} We adopt LoRA parameterized as $W = BA$, where $A \in {R}^{r \times d_{\text{in}}}$, $B \in {R}^{d_{\text{out}} \times r}$. 

\textbf{LoRA Placement.}
\textbf{ID-LoRA} ($W_{\mathrm{ID}}$) is a private, client-specific . It captures speaker timbre and channel coloration, is fine-tuned locally once, and remains \textit{permanently frozen—never uploaded or aggregated} to the server. 
In contrast, \textbf{Style-LoRA} ($W_{\mathrm{style}}$) is a \textit{federated, globally shared adapter} that modulates expressive variation and is updated collaboratively across clients.

In the \textsc{GPT-SoVITS-v4} backbone, the Style-LoRA  $W_{\mathrm{style}}$ and the speaker-specific ID-LoRA  $W_{\mathrm{ID}}$ are injected into linear projection layers. Within the GPT component, LoRA adapters are applied to all self-attention projections (\texttt{q}, \texttt{k}, \texttt{v}, \texttt{out}) as well as the two feed-forward layers (i.e., the up- and down-projection in the MLP block). For the SoVITS component, LoRA is inserted into the attention projections (\texttt{q}, \texttt{k}, \texttt{v}, \texttt{out}) within each CFM block.

\subsection{Client-Side Optimization}
Each selected client $i$ loads the frozen backbone, its private ID-LoRA $W_{\mathrm{ID}}^{i}$, and the current personalized Style-LoRA from the previous round, $W_{\mathrm{sty}}'^{i, t-1}$ (or a global model for the first round). Given a local batch $(x,y)$, we compute acoustic/semantic units $u=E(y)$.

\textbf{Timbre Cloning.} Using neutral speech samples $(x,y)$ from the client, and with a frozen speaker encoder, we update only $W_{\mathrm{ID}}^{i}$ to maximize the cosine similarity between speaker embeddings of the predicted and target waveforms. Gradients to the backbone and $W_{\mathrm{style}}$ are blocked to focus on timbre fidelity.

\textbf{Stylization.} Given expressive speech samples $(x, y)$ from a client's dataset, we perform teacher-forced decoding using the backbone and optimize only the style adapter $W_{\mathrm{style}}$ via a token-level cross-entropy loss. To ensure modularity, gradients to the ID-LoRA $W_{\mathrm{ID}}^{i}$ are blocked, isolating style learning from speaker timbre. During training, we enforce a consistent emotional style within each batch of expressive data from the same client.

Each client's model is first trained locally for $n$ steps to perform timbre cloning, followed by $m$ steps of stylization. After local adaptation, the client obtains an updated Style-LoRA  $W_{\mathrm{sty}}^{i,t}$, composed of the LoRA matrices $(A_{\mathrm{style}}^{i,t}, B_{\mathrm{style}}^{i,t})$. 
Only these style-specific parameters $W_{\mathrm{style}}^i$ in Figure~\ref{fig:fig_overview} are transmitted to the server, while the ID-LoRA $W_{\mathrm{ID}}^{i}$ remains private and stored on-device.

\subsection{Server-Side Personalized Aggregation}
To harness \textit{stylistic heterogeneity}, we perform personalized aggregation inspired by collaborative filtering in recommendation systems~\cite{isinkaye2015recommendation}. We ensure that a client can benefit most from other clients with similar speaking styles.

At round $t$, after receiving Style-LoRA updates ($A_{\mathrm{style}}^{j,t}, B_{\mathrm{style}}^{j,t}$) from $C$ clients, the server builds a personalized model $W_{style}'^{i,t}$ for each client $i$. It computes attention scores by applying a softmax to the pairwise cosine similarity of the $A$ and $B$ matrices scaled by temperature $\tau$. For client $i$, the scores $\alpha_{ij}$ are:
\[
\alpha_{ij} = \frac{\exp(\mathrm{sim}(A_{\mathrm{style}}^{i,t}, A_{\mathrm{style}}^{j,t}) / \tau)}{\sum_{k=1}^{C} \exp(\mathrm{sim}(A_{\mathrm{style}}^{i,t}, A_{\mathrm{style}}^{k,t}) / \tau)}.
\]
Inspired by prior work~\cite{huang2021personalized} on temperature-scaled aggregation in federated learning, we set $\tau$ to 0.5 to control the sharpness of the attention distribution over clients. 

To make the low-rank decomposition space more stable~\cite{zhang2024towards}, we calculate attention and aggregation for ($A_{\mathrm{style}}^{j,t}, B_{\mathrm{style}}^{j,t}$) respectively. The scores $\beta_{ij}$ are computed similarly for the $B$ matrices. These scores are used for the final aggregation:
\[
A_{\mathrm{style}}'^{i,t+1} = \sum_{j=1}^{C} \alpha_{ij} A_{\mathrm{style}}^{j,t}, \quad B_{\mathrm{style}}'^{i,t+1} = \sum_{j=1}^{C} \beta_{ij} B_{\mathrm{style}}^{j,t}.
\]
This personalized model is then sent back to the client, while the private $W_{\mathrm{ID}}$ remains on-device.

\section{Experiments}

\subsection{Setup}
\subsubsection{Datasets}
We used four publicly available datasets with emotion annotations. 
To ensure style consistency across these diverse corpora, we first pooled all datasets and then employed the emotion2vec~\cite{ma-etal-2024-emotion2vec} framework to map the various emotion labels into a unified, discrete style space. This unification process resulted in a total of 10 distinct style categories for the entire experimental setup.
Based on the original annotations, the data were also grouped into two broader categories: neutral and expressive, with details of the source datasets shown in Table \ref{tab:dataset-overview}.
The ground truth text data of the dataset is annotated by Whisper-large-v3 Turbo~\cite{radford2023robust} and has been manually corrected.

To ensure cross-lingual and cross-corpus consistency, we adopt a unified preprocessing pipeline for all speech and text. Audio is resampled to 24 kHz and stored as 16-bit PCM. Constant low-frequency hum was attenuated by applying a high-pass filter along with notch filters at 50 Hz and its harmonics. Endpoint trimming and loudness normalization are applied: leading/trailing silences are removed using a 60 ms silence threshold, and loudness is normalized in LUFS to stabilize batch statistics.

\begin{table}[h]
\centering
\caption{Statistics of datasets used in our study, including the number of \textit{heterogeneous} styles, speakers, and the train/val/test sample splits for both neutral and expressive subsets. }
\resizebox{0.48\textwidth}{!}{%
\begin{tabular}{@{}l cc l ccc@{}}
\toprule
\textbf{Dataset}  &\textbf{Styles}& \textbf{Speakers} & \textbf{Data Type} & \textbf{Train} & \textbf{Val} & \textbf{Test} \\
\midrule

\multirow{2}{*}{\textbf{ESD}} &\multirow{2}{*}{5}& \multirow{2}{*}{20}& Neutral    & 5,600& 700& 700\\
\cmidrule(l){4-7} 
 && & Expressive & 22,400& 2,800& 2,800\\
\midrule
\multirow{2}{*}{\textbf{EmoV-DB}} &\multirow{2}{*}{8}&  \multirow{2}{*}{4}& Neutral    & 1,254& 156& 158\\
\cmidrule(l){4-7} 
 && & Expressive & 4,260& 532& 533\\
\midrule
\multirow{2}{*}{\textbf{RAVDESS}} &\multirow{2}{*}{8}&  \multirow{2}{*}{24}& Neutral    & 230& 28& 30\\
\cmidrule(l){4-7} 
 && & Expressive & 921& 115& 116\\
\midrule
\multirow{2}{*}{\textbf{CREMA-D}} &\multirow{2}{*}{6}&  \multirow{2}{*}{12}& Neutral    & 869& 108& 110\\
\cmidrule(l){4-7} 
 && & Expressive & 4,355& 544& 545\\
\bottomrule
\end{tabular}%
}

\label{tab:dataset-overview}
\end{table}

\subsubsection{Baselines}
As a non-federated reference, we first include a zero-shot in-context learning (ICL) baseline using  \textsc{CosyVoice2}~\cite{du2025cosyvoice} and \textsc{GPT-SoVITS-v4}.  We also consider a fully local finetuning baseline (Local FT), where each client independently trains both ID-LoRA and Style-LoRA using private neutral and expressive data without federated sharing. To compare with existing federated learning paradigms, we include three additional baselines. FedSpeech~\cite{jiang2021fedspeech} applies selective masking to retain client-specific prosodic features, while Federated Dynamic Transformer (Fed Dy. Trans.)~\cite{hong2021federated} incrementally expands encoder and decoder layers. 
It is important to note that these established federated methods are \textbf{intrinsically coupled with their specific backbone architectures} (i.e., FastSpeech2~\cite{ren2019fastspeech} and Transformer-TTS~\cite{li2019neural}, respectively) and are \textbf{not directly compatible with LoRA techniques}. Therefore, for a fair evaluation of these methods, we adopted their original, published configurations.

\subsubsection{Training Details}
Each speaker from the datasets is treated as an independent client.
The entire training process spans 50 communication rounds, with a client participation rate of 20\% per round.
For both the private ID-LoRA and the federated Style-LoRA, we set the rank to 8 and the scaling factor to $16$. 
Local adaptation is performed using the AdamW optimizer with a batch size of $16$. We employ a learning rate of $2 \times 10^{-5}$ with a cosine decay schedule and a warmup ratio of $0.1$. For main experiments on each client, the timbre cloning phase runs for $80$ steps, followed by the stylization phase, which runs for $20$ steps. Experiments were conducted on four NVIDIA V100 GPUs.

\subsubsection{Evaluation Metrics}

We evaluate the generated speech from three perspectives: correctness, speaker similarity, and naturalness. For correctness, we report Word Error Rate (WER) and Character Error Rate (CER), calculated by comparing the Whisper-large-v3 Turbo~\cite{radford2023robust} transcription of the generated speech against the ground-truth text. Lower values indicate a more accurate and understandable synthesis.  
We evaluate Style Expressivity (SE) by applying the emotion2vec~\cite{ma-etal-2024-emotion2vec} classifier that is finetuned on the trainset to synthesized audio and use the predicted probability of the ground-truth emotion as the score, averaged over the test set. To assess speaker similarity (SS), we extract speaker embeddings using ECAPA-TDNN model ~\cite{speechbrain} and compute the cosine similarity between synthesized and reference audio samples; higher similarity scores reflect better voice cloning fidelity. Finally, to estimate perceptual naturalness, we use the nMOS metric, with blind listening evaluations conducted by 22 researchers in the laboratory, and the final evaluation results are averaged. We report cost as the cumulative upload and download bandwidth over all rounds and participating clients, counting only the trainable parameters actually transmitted. Values assume FP16 (2 bytes per parameter) and are reported in GiB.

Evaluation is conducted on the test split of each dataset (see Table~\ref{tab:dataset-overview}), comprising both neutral and expressive samples. Reported scores reflect the mean performance across all datasets.

\subsection{Main Results}
\begin{table*}[t]
\centering
\caption{Main experimental results comparing \textsc{Fed-PISA} with baseline methods and ablation studies. The best and second-best results are \textbf{bolded} and \underline{underlined}. Results are reported over 3 runs with different random seeds. }
\label{tab:main_results}
\resizebox{\textwidth}{!}{%
\begin{tabular}{@{}llrrrcccc@{}}
\toprule
\multicolumn{1}{c}{\textbf{Method}} &
\multicolumn{1}{c}{\textbf{Backbone}} &
\multicolumn{3}{c}{\textbf{Efficiency}} &
\multicolumn{2}{c}{\textbf{Correctness}} &
\multicolumn{1}{c}{\textbf{Similarity}} &
\multicolumn{1}{c}{\textbf{Naturalness}} \\
\cmidrule(lr){3-5} \cmidrule(lr){6-7} \cmidrule(lr){8-8} \cmidrule(lr){9-9}
 & & \textbf{Tuned/Total (B)}& \textbf{Round} & \textbf{Cost (GiB)} $\downarrow$ & \textbf{SE}  $\uparrow$ & \textbf{WER}  (\%) $\downarrow$& \textbf{SS}  $\uparrow$ & \textbf{nMOS}   $\uparrow$ \\
\midrule
\multicolumn{9}{l}{\textit{Zero-Shot}} \\
Foundation (ICL) & \textsc{CosyVoice2} & \small{0/0.50}& -& -& 0.659& 7.20& 0.619& 3.84\\
Foundation & \textsc{GPT-SoVITS-v4} & \small{0/0.41}& -& -& 0.605& 5.18& 0.464& 3.39\\
\midrule
\multicolumn{9}{l}{\textit{Local Personalization Baseline}} \\
Local FT (Full) & \textsc{GPT-SoVITS-v4} & \small{0.41/0.41}& -& -& 0.618& 3.12& 0.554& 3.40\\
Local FT (LoRA) & \textsc{GPT-SoVITS-v4} & \small{0.06/0.41}& -& -& 0.626& 3.35& 0.529& 3.36\\
\midrule
\multicolumn{9}{l}{\textit{Federated Baselines and Our Method}} \\
FedSpeech & \textsc{FastSpeech2-XL}& \small{0.52/0.52}& 50& \underline{145.28}& 0.416& 6.82 & 0.556& 3.77 \\
Fed Dy. Trans. & \textsc{Transformer-TTS} & \small{0.22/0.35} & 50& 456.35& 0.463 & 8.75 & 0.602 & 3.72 \\
\rowcolor{gray!20}\textbf{\textsc{Fed-PISA} (Ours)}& \textsc{GPT-SoVITS-v4} & \small{0.04/0.41}& 50& \textbf{45.8}& \textbf{0.704}& \textbf{2.70}& \textbf{0.645}& \textbf{4.08}\\
\midrule
\multicolumn{9}{l}{\textit{Ablation Studies}} \\
\textsc{Fed-PISA} w/o ID-LoRA & \textsc{GPT-SoVITS-v4} & \small{0.02/0.41} & 50 & 45.8 &  0.624& 3.02& 0.507& 3.68\\
\textsc{Fed-PISA} w/o Style-LoRA & \textsc{GPT-SoVITS-v4} & \small{0.02/0.41} & 50 &  45.8& 0.588& 3.77& 0.610& 3.55\\
FedAvg  & \textsc{GPT-SoVITS-v4} & \small{0.04/0.41} & 50 &  45.8& 0.476& 3.60& 0.523& 3.80\\
\bottomrule
\end{tabular}%
}
\end{table*}

\textbf{Comparison with Foundation and Local Baselines.} 
As presented in Table~\ref{tab:main_results}, compared to zero-shot voice cloning, \textsc{Fed-PISA} shows a dramatic improvement. This highlights the necessity of fine-tuning for high-fidelity personalization. More critically, \textsc{Fed-PISA} surpasses the Local FT (LoRA) baseline, where each client trains independently. Our collaborative approach achieves a higher SS score, which proves that \textsc{Fed-PISA}'s collaborative learning paradigm allows clients to learn richer stylistic variations from their peers, overcoming the data scarcity limitations of purely local fine-tuning.
\begin{figure}[t]
    \centering
    \includegraphics[width=1\linewidth]{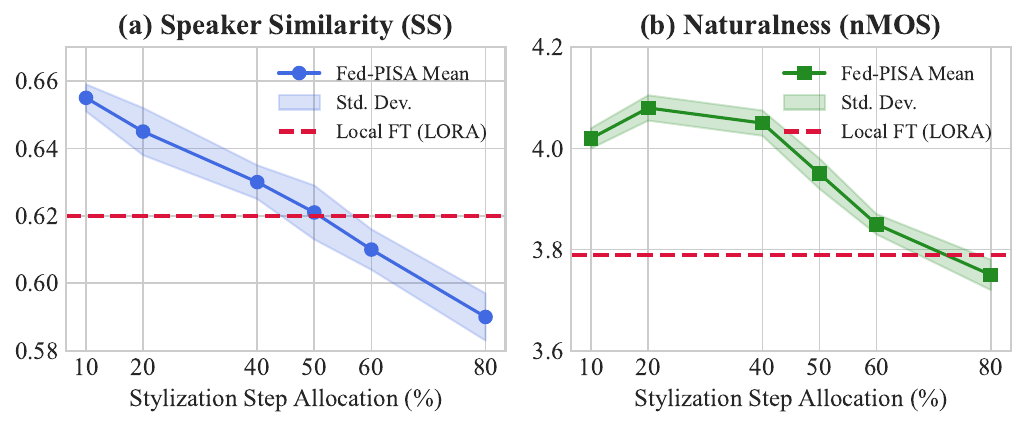}
    \caption{Performance trade-off when allocating training steps between ID-LoRA ($n$) and Style-LoRA ($m$). The solid line denotes the mean performance over three independent runs with different random seeds, and the shaded region represents the standard deviation across runs.}
    \label{fig:ablation_steps}
\end{figure}

\noindent\textbf{Comparison with Federated Baselines.}
Existing federated baselines exhibit even lower style expressivity than zero-shot voice cloning methods, suggesting that prior FL approaches tend to suppress the stylistic heterogeneity that \textsc{Fed-PISA} effectively preserves and leverages. In addition, \textsc{Fed-PISA} achieves substantially better performance in terms of WER, SS, and SpeechMOS. This highlights a key advantage of our paradigm: its flexibility and applicability to modern, more powerful foundation models. The results in Table \ref{tab:main_results} demonstrate that the Fed-PISA framework, by effectively leveraging such models via PEFT~\cite{bondaruk2025lorp, kwon2025parameter, fujita2024lightweight}, significantly outperforms these legacy federated systems.

\noindent\textbf{Efficiency Analysis.}
\textsc{Fed-PISA} demonstrates strong efficiency advantages. Owing to the use of LoRA, its trainable parameter count is approximately 1/5 that of Fed Dy. Trans. under comparable backbone sizes. In terms of communication cost, \textsc{Fed-PISA} incurs only 1/10 the costs of Fed Dy. Trans., and about 1/3 that of FedSpeech, despite the latter employing a masking mechanism to reduce transmission load. 

\subsection{Ablation Studies}
Ablation studies (Table \ref{tab:main_results}) validate our key design choices.

\noindent\textbf{Identity-Style Disentanglement:} The ablation results confirm the necessity of our identity-style disentanglement, as removing either the private ID-LoRA or the collaborative Style-LoRA leads to a significant degradation in speaker similarity and naturalness. This is because a single adapter struggles to reconcile the competing objectives of preserving a private, static identity and learning from diverse, shared styles, whereas our disentangled design allows both goals to be optimized without mutual interference.

\noindent\textbf{Personalized Aggregation:} The comparison with FedAvg~\cite{zhang2024towards} is most telling, as it is conducted under an identical backbone and LoRA setup, providing a perfectly controlled ablation of our aggregation strategy. It demonstrates that a naive aggregation of style updates leads to a ``style-averaging" effect that harms both speaker identity and expressive quality. In contrast, our attention-based personalized aggregation successfully leverages stylistic similarities between clients with \textit{heterogeneous styles}.

\subsection{Analysis on Training Step Allocation}
\label{sec:analysis_steps}
We analyzed the trade-off between timbre cloning ($n$ steps) and stylization ($m$ steps) on the held-out validation by varying their allocation while keeping the total steps constant at 100. We found that speaker similarity (SS) monotonically decreased with increasing stylization, while naturalness peaked at a 20\% allocation ($m$ = 20). Excessive stylization corrupted speaker identity and degraded naturalness. This validates our choice of $n=80$ and $m=20$, which strikes an optimal balance without consulting the test set.


\section{Conclusion}
We introduced \textsc{FED-PISA} to address the central challenge in federated voice cloning: leveraging stylistic heterogeneity across clients without incurring high communication costs or degrading speaker identity. Our framework resolves this by disentangling a private ID-LoRA from a collaborative Style-LoRA and employing a personalized aggregation strategy. Experiments show significant improvements in style expressivity and speaker similarity over federated baselines.

\section{Compliance with Ethical Standards}
Compliance with Ethical Standards. This study was performed in line with the principles of the Declaration of Helsinki. Approval was granted by the Ethics Committee of the Institute of Computing Technology, Chinese Academy of Sciences (ICT, CAS). All participants in the blind listening tests provided informed consent. Only anonymized ratings were collected. We additionally used open-access human-speech datasets strictly in accordance with their licenses. No new recordings were collected from these corpora.
\bibliographystyle{IEEEbib}
\bibliography{strings,refs}

\begin{thebibliography}{10}

\bibitem{du2025cosyvoice}
Zhihao Du, Yuxuan Wang, Qian Chen, Xian Shi, Xiang Lv, Tianyu Zhao, Zhifu Gao, Yexin Yang, Changfeng Gao, Hui Wang, et~al.,
\newblock ``Cosyvoice 2: Scalable streaming speech synthesis with large language models,''
\newblock {\em arXiv preprint arXiv:2412.10117}, 2024.

\bibitem{le2023voicebox}
Matthew Le, Apoorv Vyas, Bowen Shi, Brian Karrer, Leda Sari, Rashel Moritz, Mary Williamson, Vimal Manohar, Yossi Adi, Jay Mahadeokar, et~al.,
\newblock ``Voicebox: Text-guided multilingual universal speech generation at scale,''
\newblock {\em Advances in neural information processing systems}, vol. 36, pp. 14005--14034, 2023.

\bibitem{xie2025fireredtts}
Kun Xie, Feiyu Shen, Junjie Li, Fenglong Xie, Xu~Tang, and Yao Hu,
\newblock ``Fireredtts-2: Towards long conversational speech generation for podcast and chatbot,''
\newblock {\em arXiv e-prints}, pp. arXiv--2509, 2025.

\bibitem{meng2025ds}
Ming Meng, Ziyi Yang, Jian Yang, Zhenjie Su, Yonggui Zhu, and Zhaoxin Fan,
\newblock ``Ds-tts: Zero-shot speaker style adaptation from voice clips via dynamic dual-style feature modulation,''
\newblock {\em arXiv preprint arXiv:2506.01020}, 2025.

\bibitem{chen2024emoknob}
Haozhe Chen, Run Chen, and Julia Hirschberg,
\newblock ``Emoknob: Enhance voice cloning with fine-grained emotion control,''
\newblock in {\em Proceedings of the 2024 Conference on Empirical Methods in Natural Language Processing}, 2024, pp. 8170--8180.

\bibitem{huang2023personalized}
Sung-Feng Huang, Chia-Ping Chen, Zhi-Sheng Chen, Yu-Pao Tsai, and Hung-yi Lee,
\newblock ``Personalized lightweight text-to-speech: Voice cloning with adaptive structured pruning,''
\newblock in {\em ICASSP 2023-2023 IEEE International Conference on Acoustics, Speech and Signal Processing (ICASSP)}. IEEE, 2023, pp. 1--5.

\bibitem{comini2025lightweight}
Giulia Comini, Heereen Shim, and Manuel~Sam Ribeiro,
\newblock ``Lightweight neural front-ends for low-resource on-device text-to-speech,''
\newblock in {\em ICASSP 2025-2025 IEEE International Conference on Acoustics, Speech and Signal Processing (ICASSP)}. IEEE, 2025, pp. 1--5.

\bibitem{bondaruk2025lorp}
{\L}ukasz Bondaruk and Jakub Kubiak,
\newblock ``Lorp-tts: Low-rank personalized text-to-speech,''
\newblock {\em arXiv preprint arXiv:2502.07562}, 2025.

\bibitem{kwon2025parameter}
Ki-Joong Kwon, Jun-Ho So, and Sang-Hoon Lee,
\newblock ``Parameter-efficient fine-tuning for low-resource text-to-speech via cross-lingual continual learning,''
\newblock in {\em Proc. Interspeech 2025}, 2025, pp. 1613--1617.

\bibitem{fujita2024lightweight}
Kenichi Fujita, Takanori Ashihara, Marc Delcroix, and Yusuke Ijima,
\newblock ``Lightweight zero-shot text-to-speech with mixture of adapters,''
\newblock in {\em Proc. Interspeech 2024}, 2024, pp. 692--696.

\bibitem{zhang2021survey}
Chen Zhang, Yu~Xie, Hang Bai, Bin Yu, Weihong Li, and Yuan Gao,
\newblock ``A survey on federated learning,''
\newblock {\em Knowledge-Based Systems}, vol. 216, pp. 106775, 2021.

\bibitem{jiang2021fedspeech}
Ziyue Jiang, Yi~Ren, Ming Lei, and Zhou Zhao,
\newblock ``Fedspeech: Federated text-to-speech with continual learning,''
\newblock in {\em Proceedings of the Thirtieth International Joint Conference on Artificial Intelligence (IJCAI-21)}, 2021, pp. 3829--3835.

\bibitem{hong2021federated}
Zhenhou Hong, Jianzong Wang, Xiaoyang Qu, Jie Liu, Chendong Zhao, and Jing Xiao,
\newblock ``Federated learning with dynamic transformer for text to speech,''
\newblock in {\em Proc. Interspeech 2021}, 2021, pp. 3590--3594.

\bibitem{lou2024stylespeech}
Haowei Lou, Hye-Young Paik, Wen Hu, and Lina Yao,
\newblock ``Stylespeech: Parameter-efficient fine tuning for pre-trained controllable text-to-speech,''
\newblock in {\em Proceedings of the 6th ACM International Conference on Multimedia in Asia}, 2024, pp. 1--7.

\bibitem{isinkaye2015recommendation}
Folasade~Olubusola Isinkaye, Yetunde~O Folajimi, and Bolande~Adefowoke Ojokoh,
\newblock ``Recommendation systems: Principles, methods and evaluation,''
\newblock {\em Egyptian informatics journal}, vol. 16, no. 3, pp. 261--273, 2015.

\bibitem{huang2021personalized}
Yutao Huang, Lingyang Chu, Zirui Zhou, Lanjun Wang, Jiangchuan Liu, Jian Pei, and Yong Zhang,
\newblock ``Personalized cross-silo federated learning on non-iid data,''
\newblock in {\em Proceedings of the AAAI conference on artificial intelligence}, 2021, vol.~35, pp. 7865--7873.

\bibitem{zhang2024towards}
Jianyi Zhang, Saeed Vahidian, Martin Kuo, Chunyuan Li, Ruiyi Zhang, Tong Yu, Guoyin Wang, and Yiran Chen,
\newblock ``Towards building the federatedgpt: Federated instruction tuning,''
\newblock in {\em ICASSP 2024-2024 IEEE International Conference on Acoustics, Speech and Signal Processing (ICASSP)}. IEEE, 2024, pp. 6915--6919.

\bibitem{ma-etal-2024-emotion2vec}
Ziyang Ma, Zhisheng Zheng, Jiaxin Ye, Jinchao Li, Zhifu Gao, ShiLiang Zhang, and Xie Chen,
\newblock ``emotion2vec: Self-supervised pre-training for speech emotion representation,''
\newblock in {\em Findings of the Association for Computational Linguistics: ACL 2024}, Lun-Wei Ku, Andre Martins, and Vivek Srikumar, Eds., Bangkok, Thailand, Aug. 2024, pp. 15747--15760, Association for Computational Linguistics.

\bibitem{radford2023robust}
Alec Radford, Jong~Wook Kim, Tao Xu, Greg Brockman, Christine McLeavey, and Ilya Sutskever,
\newblock ``Robust speech recognition via large-scale weak supervision,''
\newblock in {\em International conference on machine learning}. PMLR, 2023, pp. 28492--28518.

\bibitem{ren2019fastspeech}
Yi~Ren, Yangjun Ruan, Xu~Tan, Tao Qin, Sheng Zhao, Zhou Zhao, and Tie-Yan Liu,
\newblock ``Fastspeech: Fast, robust and controllable text to speech,''
\newblock {\em Advances in neural information processing systems}, vol. 32, 2019.

\bibitem{li2019neural}
Naihan Li, Shujie Liu, Yanqing Liu, Sheng Zhao, and Ming Liu,
\newblock ``Neural speech synthesis with transformer network,''
\newblock in {\em Proceedings of the AAAI conference on artificial intelligence}, 2019, vol.~33, pp. 6706--6713.

\bibitem{speechbrain}
Mirco Ravanelli, Titouan Parcollet, Peter Plantinga, Aku Rouhe, Samuele Cornell, Loren Lugosch, Cem Subakan, Nauman Dawalatabad, Abdelwahab Heba, Jianyuan Zhong, Ju-Chieh Chou, Sung-Lin Yeh, Szu-Wei Fu, Chien-Feng Liao, Elena Rastorgueva, François Grondin, William Aris, Hwidong Na, Yan Gao, Renato~De Mori, and Yoshua Bengio,
\newblock ``{SpeechBrain}: A general-purpose speech toolkit,'' 2021,
\newblock arXiv:2106.04624.

\end{thebibliography}
\end{document}